\documentclass[aps,prl,twocolumn,superscriptaddress,showpacs,floatfix]{revtex4-2}

\usepackage{graphicx}
\usepackage{dcolumn}
\usepackage{bm}
\usepackage{amsmath}
\usepackage{hyperref}
\usepackage{amssymb}
\usepackage{xcolor}

\begin{document}

\title{A Question of Shape: New Mechanism Governing Superheavy Nuclei Survival}

\author{A. Rahmatinejad}
\affiliation{Joint Institute for Nuclear Research, Dubna, 141980, Russia}
\author{T. M. Shneidman}
\affiliation{Joint Institute for Nuclear Research, Dubna, 141980, Russia}
\author{G. G. Adamian}
\affiliation{Joint Institute for Nuclear Research, Dubna, 141980, Russia}
\author{N. V. Antonenko}
\affiliation{Joint Institute for Nuclear Research, Dubna, 141980, Russia}
\author{P.~Jachimowicz}
 \affiliation{Institute of Physics,
University of Zielona G\'{o}ra, Szafrana 4a, 65516 Zielona G\'{o}ra, Poland}

\author{M.~Kowal} \email{michal.kowal@ncbj.gov.pl}
\affiliation{National Centre for Nuclear Research, Pasteura 7, 02-093 Warsaw, Poland}

\begin{abstract}
We demonstrate that hot superheavy nuclei do not retain spherical shapes, as traditionally assumed, but instead equilibrate in deformed—often oblate or triaxial—configurations at finite excitation energy. This behavior arises from a mechanism analogous to the Jahn--Teller effect: spherical systems exhibit high single-particle degeneracy near the Fermi surface, causing their shell corrections to damp out significantly faster with temperature than those of deformed shapes. Using a finite-temperature framework, we reveal a thermally induced inversion of the potential-energy landscape in the $Z=118$--$120$ region, where deformed minima become energetically favored at $U \approx 30$--$50$~MeV. This shape inversion fundamentally alters the competition between neutron evaporation and fission. We derive a deformation-dependent correction to the survival probability, revealing a systematic bias in estimates based on spherical ground-state properties.
Our results identify a finite-temperature structure effect that calls for a revision of current models of superheavy nucleus synthesis and decay.
\end{abstract}


\maketitle

\paragraph{Introduction.---}
The laboratory synthesis of superheavy nuclei (SHN) stands as a monumental achievement, fundamentally extending the limits of the periodic table \cite{OGRYK}. These systems, hosting the strongest Coulomb fields in nature, represent the ultimate frontier of bound matter. In this regime, where immense electrostatic repulsion nearly cancels the nuclear surface tension, the macroscopic liquid-drop energy is remarkably flat. Consequently, the very existence of SHN is determined primarily by quantum-mechanical shell effects—oscillatory corrections arising from the "bunching" of single-particle states into stabilizing shells separated by energy gaps.

Crucially, however, nuclei with $Z \ge 114$ are synthesized exclusively via hot fusion \cite{Oganessian2000,Oganessian2006,OganessianUtyonkov2015_RPP,OganessianUtyonkov2015_NPA,Hofmann2007,Hofmann2012,Dullmann2010}, producing compound nuclei with substantial excitation energies ($U \approx 30-40$ MeV). This thermal regime fundamentally challenges the standard static description: excitation quenches shell effects, erodes fission barriers, and modifies the potential energy surface (PES). The question of survival probability in this dynamic, is therefore the critical factor governing synthesis success.

In this Letter, we identify a universal microscopic mechanism governing this thermal evolution—a nuclear analogue of the Jahn-Teller effect \cite{Jahn1937, Englman1972,LL_QM_1977,Khomskii2014, Wilkinson2002}. We demonstrate that the high spectral degeneracy of spherical shapes, while providing essential stabilization at zero temperature, renders them remarkably fragile against heating. Excitation energy acts as a symmetry-breaking perturbation, driving the nucleus away from the highly degenerate spherical shape toward deformed configurations, where the lower spectral density ensures more robust shell correlations.
\begin{figure}[t]
    \centering
    \includegraphics[width=1\linewidth]{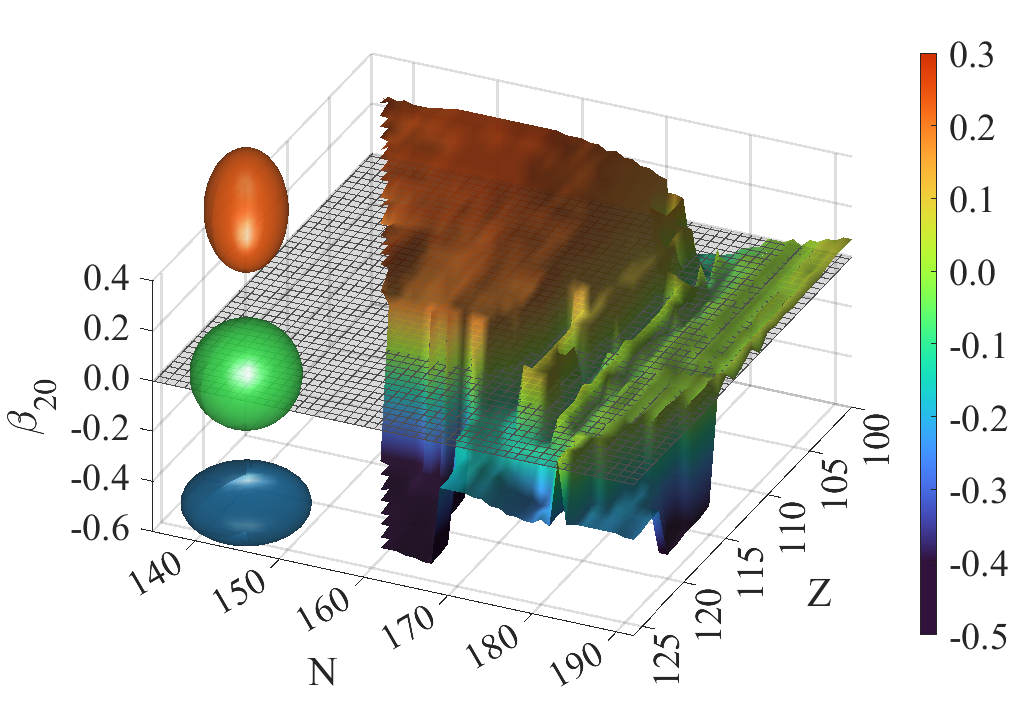}
    \caption{Landscape of ground-state quadrupole deformations $\beta_{20}$ for superheavy nuclei.
    The data are adapted from the macroscopic-microscopic calculations in Ref.~\cite{Tables}.}
\label{b23D}
\end{figure}

Figure~\ref{b23D} maps the ground-state quadrupole deformation $\beta_{20}$ for cold SHN, calculated via our macroscopic-microscopic approach assuming axial symmetry \cite{Tables}. The landscape is topologically complex, featuring spherical, well-deformed prolate, oblate, and superdeformed oblate minima \cite{SDO1,SDO2}.  Notably, the heaviest systems ($Z \geqslant 118$) exhibit a vibrant diversity of equilibrium configurations.

\paragraph{Thermodynamics of hot nucleus.---}
The compound nucleus formed in a heavy-ion reaction is treated as an isolated system, and its global evolution is constrained by strict energy conservation. For any given deformation $\beta$, the total energy $E_{0}$ is fixed and can be written as
\begin{equation}
E_{0} = U + E_\mathrm{pot}^\mathrm{GS} = U^{*}(\beta) + E_\mathrm{pot}(U^{*},\beta),
\end{equation}
where $U$ denotes the total energy measured with respect to the ground-state potential energy $E_\mathrm{pot}^\mathrm{GS}$, and $U^{*}(\beta)$ is the excitation energy available at deformation $\beta$. The effective potential energy is decomposed into macroscopic and microscopic contributions,
\begin{equation}\label{eq20b}
E_\mathrm{pot}(U^{*},\beta) = E_\mathrm{mac}(\beta) + E_\mathrm{mic}(U^{*},\beta),
\end{equation}
with the macroscopic liquid-drop part $E_\mathrm{mac}(\beta)$ assumed to be independent of excitation energy, while the microscopic component $E_\mathrm{mic}(U^{*},\beta)$ reflects the thermal damping of shell and pairing correlations.

During collective shape evolution, energy is continuously redistributed between collective and intrinsic degrees of freedom, and the system must be described microcanonically. However, the collective shape coordinates evolve on a much slower timescale than the intrinsic single-particle motion, which justifies an adiabatic approximation for the collective dynamics. The evolution can therefore be viewed as a coarse-graining in deformation space: at each fixed deformation $\beta$ the collective variables are effectively frozen, while the intrinsic degrees of freedom rapidly equilibrate among themselves. Each such configuration is thus described by a local canonical ensemble, characterized by a deformation-dependent temperature $T_{\beta}$ determined self-consistently by energy conservation through
\begin{eqnarray}
U(T_{\beta}) + E_\mathrm{pot}(T_{\beta},\beta) = E_{0}, \qquad U(T_{\beta}) = U^{*}(\beta).
\end{eqnarray}

The assumption of local canonical equilibrium at fixed deformation allows the microscopic energy correction to be expressed as the correction to the free energy,
\begin{equation}\label{eq20bb}
E_\mathrm{mic}(U^{*},\beta) = \delta F_\mathrm{shell}(T_{\beta}) + \delta F_\mathrm{pair}(T_{\beta}).
\end{equation}
Since the available excitation energy varies with deformation, the temperature $T_{\beta}$ is a local quantity and no global free-energy surface is defined. Instead, $E_\mathrm{pot}(U^{*},\beta)$ represents an effective potential-energy surface constructed under the assumption that the damping of shell and pairing correlations occurs at the local temperature dictated by the microcanonical energy constraint.

\paragraph{Thermally induced shape competition.---}
As is well established, highly excited nuclei typically transmit from deformed to spherical shapes. This standard "shape restoration" occurs because the macroscopic liquid-drop potential usually favors sphericity ($\beta=0$), dominating once the deformation-driving shell effects are thermally quenched. However, in SHN, this paradigm is inverted. Here, the macroscopic potential is flat or unstable to deformation due to the strong Coulomb repulsion. Consequently, the thermal evolution of the shape is not driven by a macroscopic restoring force, but rather by the differential fading of shell effects themselves.
\begin{figure*}[]
\begin{center}
\includegraphics[width=0.32\linewidth]{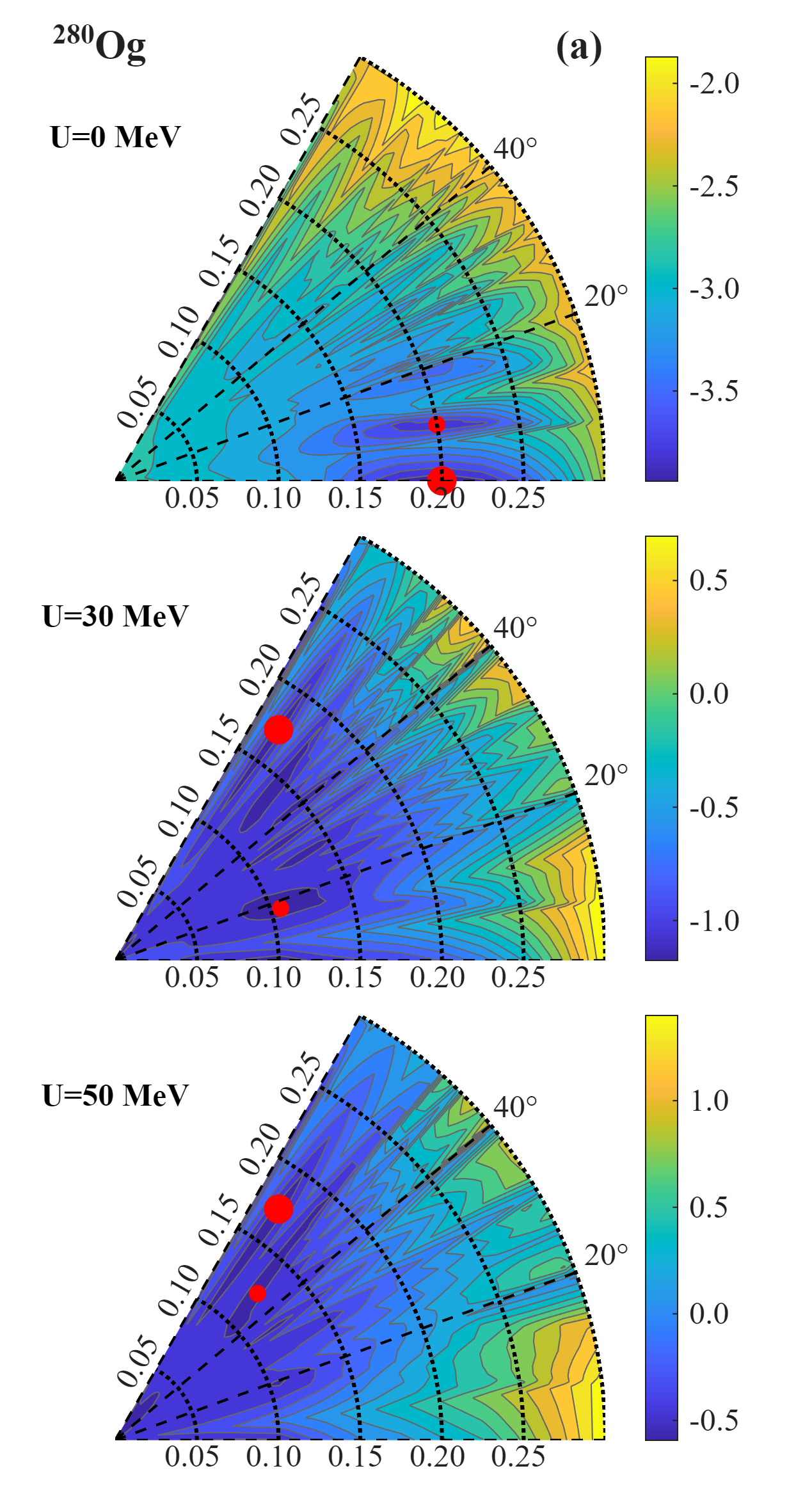}
\includegraphics[width=0.32\linewidth]{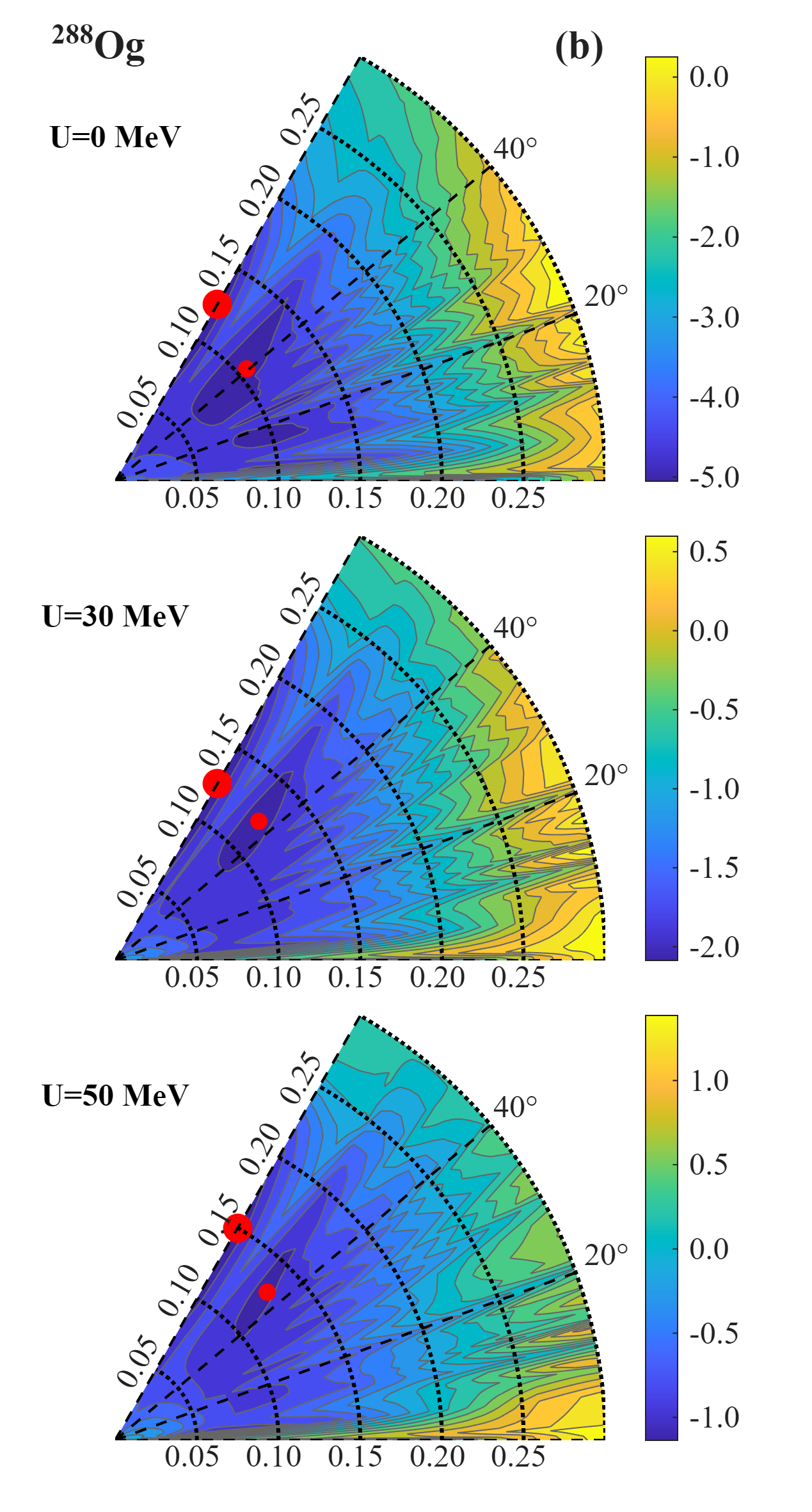}
\includegraphics[width=0.32\linewidth]{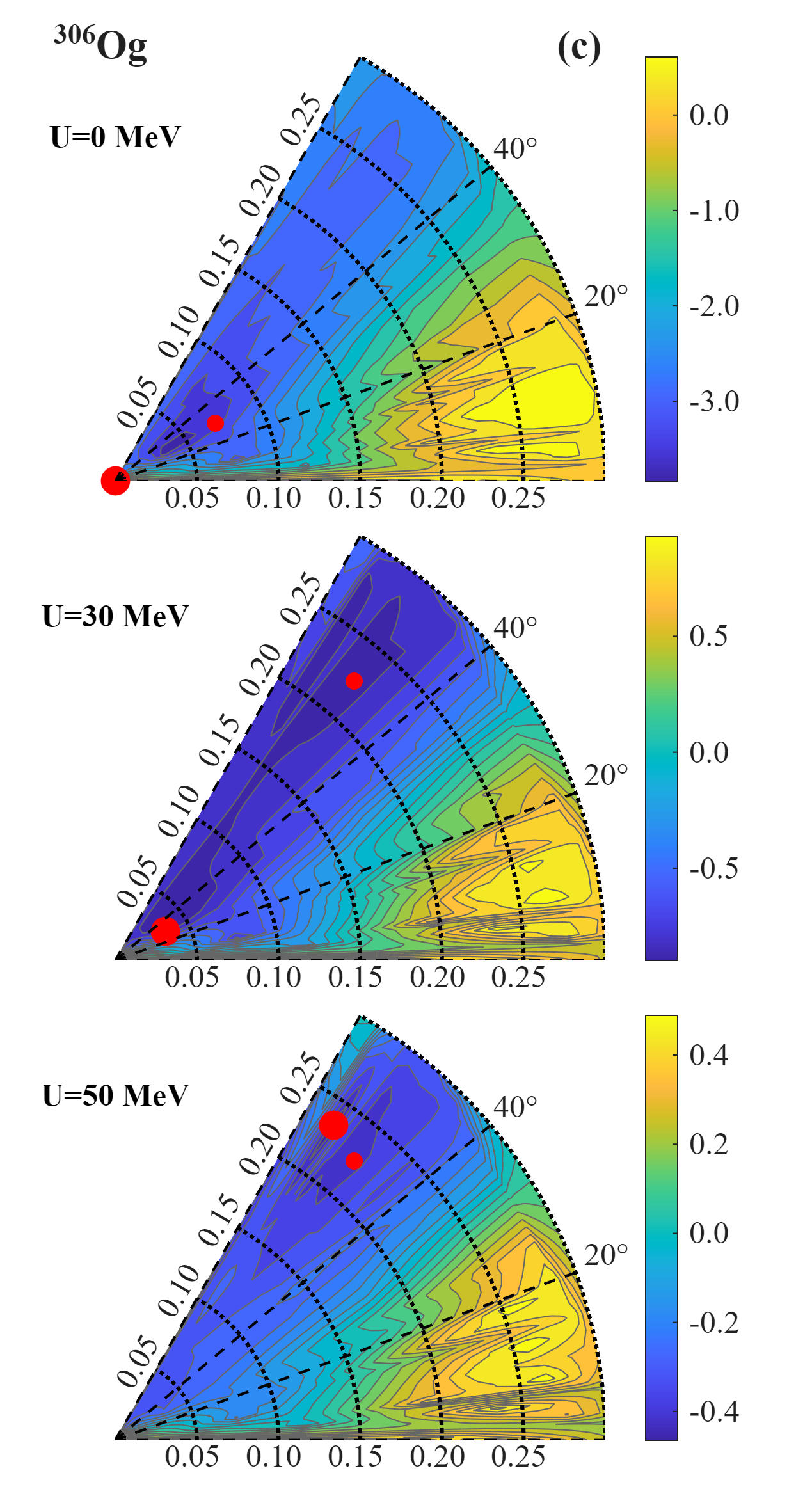}
\caption{Evolution of the potential energy surface (PES) in the plane of $(\beta, \gamma)$ for three isotopes of element $Z=118$: $^{280}$Og (a), $^{288}$Og (b), and $^{306}$Og (c). The panels display the energy landscapes at three different excitation energies relative to the ground state: $U=0$ (top), $30$ (middle), and $50$ MeV (bottom), with contour lines spaced by 1 MeV. The large red dot marks the global minimum, while the small red dot indicates a competing local minimum.}
\label{3plots}
\end{center}
\end{figure*}
To elucidate this mechanism, we track the thermal evolution of PES minima for $^{280}$Og, $^{288}$Og, and $^{306}$Og (Fig.~\ref{3plots}). Calculations were performed on a $(\beta_{20}, \beta_{22})$ grid (step 0.025) up to deformation 0.25. Using Eqs.~(\ref{eq20b}-\ref{eq20bb}), we determined the effective potential energy at total excitation energies of $U=30$ and $50$ MeV relative to the ground state (further details on these calculations can be found in Ref. Ref.~\cite{Rahmatinejad2024}).
The global minimum is marked by a large red circle and the competing one by a small red circle.

For the neutron-deficient isotope $^{280}$Og (Fig.~\ref{3plots} (a)), the ground state is a well-deformed prolate configuration. As excitation energy increases, a distinct shape evolution emerges: after $U=30$ MeV, the minimum settles in a well-deformed oblate configuration.
This prolate-to-oblate transition highlights the dynamic nature of the heated potential landscape.

In contrast, $^{288}$Og (Fig.~\ref{3plots} (b)) displays remarkable thermal stability. The deformed minimum at $(\beta_{20}, \beta_{22}) \approx (0.10, 0.10)$ remains fixed even as excitation increases to $U=50$ MeV. Here, the macroscopic preference for deformation aligns with the microscopic shell effects, suppressing any shape transition (an "oblate-to-oblate" stability).

Finally, the most dramatic change occurs in the heavy isotope $^{306}$Og (Fig.~\ref{3plots} (c)), situated near the spherical $N=184$ shell closure. The ground state ($U=0$) is spherical $(\beta_{20}, \beta_{22}) \approx (0, 0)$. However, already at $U=30$ MeV, the rapid damping of the spherical shell correction destabilizes this shape. The minimum abruptly jumps to a well-deformed oblate/triaxial configuration, persisting with minor fluctuations up to $U=50$ MeV. This spherical-to-oblate transition confirms that for spherical SHN, high excitation acts as a driver for spontaneous deformation.
As seen from Fig.~\ref{3plots}, similar to the global minima, competing minima also evolve towards an oblate configuration by increasing their triaxiality with excitation energy.

These results underscore the pivotal role of deformation-dependent shell damping. In SHN where the macroscopic potential is essentially flat and lacks a stabilizing minimum at sphericity ($\beta_{20}=0$), the rapid washing out of spherical shell corrections naturally favors deformed configurations at high excitation. This thermally induced shape transition constitutes a new, microscopic mechanism driving the competition between equilibrium shapes in hot superheavy nuclei.

\paragraph{Rate of symmetry breaking.---}
We assemble the single-particle levels into quasi-degenerate groups with a width of $\delta E = 0.2$ MeV (results are not sensitive to the specific choice of $\delta E$). We analyze the level distribution within an energy window of $\pm 5$ MeV around the Fermi surface $\varepsilon_{f}$.
Levels with $\varepsilon \leqslant \varepsilon_{f}$ form quasi-degenerate groups of occupied states, while those with $\varepsilon > \varepsilon_{f}$ constitute the unoccupied groups available for thermal excitation.

\begin{figure}[]
\begin{center}
\includegraphics[width=0.45\textwidth] {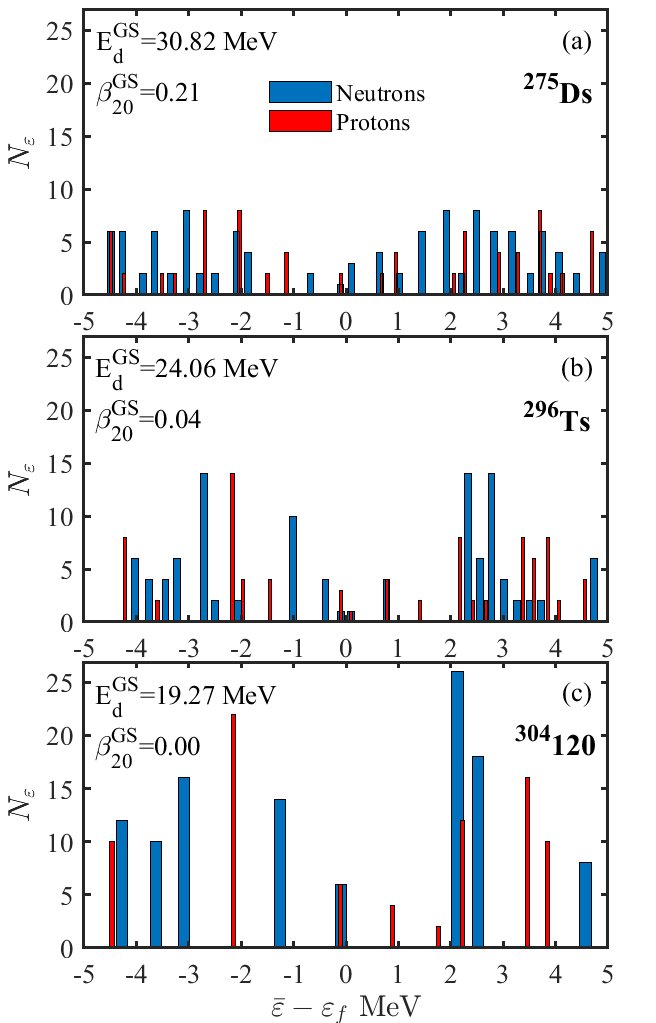}
\caption{Single-particle level degeneracy near the Fermi surface ($\varepsilon_{f} \pm 5$ MeV). The number of states per quasi-degenerate group, $N_{\varepsilon}$, is shown vs. mean group energy $\bar{\varepsilon} - \varepsilon_{f}$ for (a) deformed $^{275}$Ds ($\beta_{20}^\mathrm{GS}=0.21$), (b) transitional $^{296}$Ts ($\beta_{20}^\mathrm{GS}=0.04$), and (c) spherical $^{304}$120 ($\beta_{20}^\mathrm{GS}=0$) nuclei.}
\label{Num3Plot}
\end{center}
\end{figure}

In Fig.~\ref{Num3Plot}, we present the number of single-particle levels for each quasi-degenerate group around Fermi energy, plotted against the mean energy value for those groups.
The calculations utilize neutron and proton single-particle spectra for three nuclei representative of distinct deformation regimes: the well-deformed $^{275}$Ds ($\beta_{20}^\mathrm{GS}=0.21$), the weakly deformed $^{296}$Ts ($\beta_{20}^\mathrm{GS}=0.04$), and the spherical $^{304}$120 ($\beta_{20}^\mathrm{GS}=0$).
The damping parameter
\begin{eqnarray} E_{d}^{\mathrm{GS}}=\left(\frac{\mathrm{d}\log(E_\mathrm{mic}(U^*,\beta^\mathrm{GS}))}{\mathrm{d}U^*}\right)^{-1}
\end{eqnarray}
averaged in the range of $U^*=10-50$ MeV is used to evaluate the damping rates at $\beta=\beta^\mathrm{GS}$.

In $^{275}$Ds (Fig.~\ref{Num3Plot}~(a)), the quasi-degenerate groups are approximately evenly spaced with a notably low density of states per group. This sparse distribution implies that thermal redistribution of nucleons requires overcoming significant energy gaps between these small groups. Consequently, the shell structure remains robust against thermal fluctuations, leading to a strong persistence of shell effects with increasing temperature ($E_{d}^\mathrm{GS}=30.82$ MeV).

In the intermediate case of $^{296}$Ts (Fig.~\ref{Num3Plot}~(b)), the landscape changes. While the density of states remains low near the Fermi surface, a notable increase in degeneracy appears in the unoccupied states above 2 MeV. This enhanced availability of excited states facilitates thermal transitions, leading to a more rapid fading of shell corrections than in the well-deformed case ($E_{d}^\mathrm{GS}=24.06$ MeV).

Finally, the spherical nucleus $^{304}$120 (Fig.~\ref{Num3Plot}~(c)) exhibits the most dramatic structure. Here, the spectrum is dominated by two highly degenerate groups of states near the Fermi level. This high degeneracy makes a vast number of configurations accessible even at very low excitation energies, driving an exceptionally rapid damping of shell effects ($E_{d}^\mathrm{GS}=19.27$ MeV).

A secondary factor influencing the damping rate is the energy gap to the lowest excited state. In spherical nuclei, the high degeneracy often leads to a large energy gap separating the ground state from the first excited state. Conversely, in deformed nuclei, the lifting of degeneracy typically creates a denser spectrum with lower-lying excited states. Naively, one might expect this to facilitate damping in deformed systems.
However, when comparing the ground states of different SHN with similar magnitudes of shell corrections, the gap between the Fermi surface and the first excited states is generally comparable. Under these conditions, the determining factor becomes the \textit{density} of the available states (degeneracy) rather than the distance to the first state. Consequently, the high degeneracy of spherical shells dominates, driving the rapid damping described above. While the simultaneous contributions of the neutron and proton subsystems add complexity, the overarching trend remains: high degeneracy accelerates the thermal destruction of shell correlations.

A distinct scenario arises when the Fermi level lies within an open shell, effectively placing it within a highly degenerate group of single-particle states. In such a configuration, the high density of accessible states at the Fermi surface would lead to an immediate and rapid damping of microscopic corrections, even at negligible excitation energies. However, this scenario is physically atypical for the minima of PES of SHN.
The "open shell" case with high degeneracy at the Fermi level contradicts the condition of stability required for a heavy nucleus to exist in the first place.

\begin{figure}[]
\begin{center}
\includegraphics[width=0.45\textwidth] {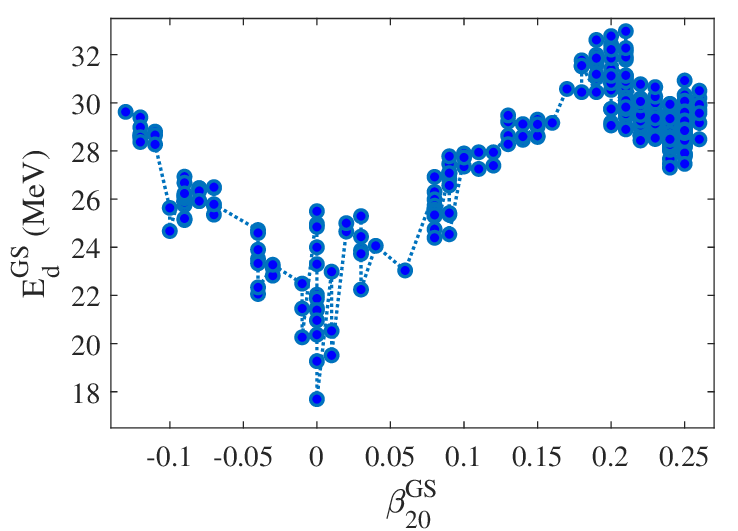}
\caption{Damping parameters $E_{d}^\mathrm{GS}$ of the GS-microscopic corrections for 353 isotopes of nuclei with $Z=98-120$ plotted as a function of their quadrupole deformations.}
\label{Betadependence}
\end{center}
\end{figure}

To illustrate the deformation dependence of shell effects damping, we perform calculations for the 353 nuclei in $Z=98-120$ region (Fig.~\ref{Betadependence}). The damping parameter has large and almost constant value $E_{d}^\mathrm{GS}\approx30$ MeV for $|\beta_{20}^\mathrm{GS}|>0.15$ (shell effects survive longer), and strongly varies from 17 up to 30 for near spherical deformations.

The persistence of structural effects at high excitation energies in deformed nuclei, in contrast to their fragility in some spherical systems, offers a novel criterion for identifying the actual physical ground state on the potential energy surface. This is particularly relevant in the "flat" landscapes typical of SHN, where spherical and deformed minima often coexist with nearly identical energies. Our results suggest that thermal stability should be considered alongside static energy minimization.

\paragraph{Implications for survival probabilities.---}
These findings fundamentally reshape the theoretical framework for the survival of excited SHN. The PES maps (Fig.~\ref{3plots}) reveal a dramatic thermal erosion of the fission barrier, which would conventionally imply a collapse of survival probabilities. Here, we encounter our first surprise: strict energy conservation introduces a critical mitigating factor — the energy cost of shell damping. Since a portion of the excitation energy is consumed to ``wash out'' shell correlations, the effective thermal energy available to drive fission is reduced.

To quantify this interplay, we employ the Bohr-Wheeler formalism in the static approximation, assuming fixed deformations for the ground state ($\beta^\mathrm{GS}$) and saddle point ($\beta^\mathrm{SP}$) and neglecting minor thermal fluctuations at the saddle. Under these conditions, the ratio of neutron emission width $\Gamma_n$ to fission width $\Gamma_{f}$ is governed by the ratio of available level densities:
\begin{equation}\label{eq7gngf}
\frac{\Gamma_{n}}{\Gamma_{f}}(U^*) \propto
\frac{\rho(A-1,\beta^\mathrm{GS}, E_{0} - E_\mathrm{pot}(A-1,U^*) - B_n)}
{\rho(A,\beta^\mathrm{SP}, E_{0} - E_\mathrm{pot}^\mathrm{SP}(A))},
\end{equation}
where $B_n$ is the neutron separation energy. Crucially, the effective potential energy of the daughter nucleus, $E_\mathrm{pot}(A-1,U^*)$, must account for the shell effects damping:
\begin{equation}
E_\mathrm{pot}(A-1,U^*) = E_\mathrm{pot}^\mathrm{GS}(A-1) + \delta E_{\mathrm{mic}}(A-1,U^*).
\end{equation}
Here, $\delta E_{\text{mic}}$ represents the "latent energy" consumed by the damping process:
\begin{equation}
\delta E_{\text{mic}}(A-1,U^*) = E_{\text{mic}}(A-1,\beta^\mathrm{GS},U^*) - E_{\text{mic}}^\mathrm{GS}(A-1).
\end{equation}

Expanding the logarithm of the level density of the daughter nucleus with respect to the small parameter $\delta E_{\text{mic}}(A-1)$, we can rewrite Eq.~\eqref{eq7gngf} as:
\begin{equation}\label{eqHind}
\frac{\Gamma_{n}}{\Gamma_{f}}(U^*) = \left(\frac{\Gamma_{n}}{\Gamma_{f}}\right)_0 e^{-\delta E_{\text{mic}}(A-1,U^*)/T},
\end{equation}
where $T = (\partial \ln \rho / \partial E)^{-1}$ corresponds to the microcanonical nuclear temperature. Here, the prefactor $(\Gamma_{n}/\Gamma_{f})_0$ represents the standard width ratio calculated without accounting for the thermal damping of shell effects.

Crucially, Eq.~\eqref{eqHind} demonstrates that the suppression of the survival probability is driven by the reduction of the microscopic correction in the daughter nucleus (the exponential term), rather than by a direct modification of the fission barrier height itself. Here we encounter another, previously unrecognized surprise: the dominant mechanism behind the hindrance is not the barrier erosion per se, but the diminished shell stabilization of the evaporation residue.

Taking damping parameter $E_d^{\mathrm{GS}}\approx 30$ MeV (see Fig.~\ref{Betadependence}), and $T\approx 1$ MeV, the reduction factor from Eq.~\eqref{eqHind} for a typical microscopic correction of $-6$ MeV is of the order of 0.02-0.01 for the energy range of 30-50 MeV.

We now introduce a completely new ingredient into estimates of the survival probability of hot nuclei. Nammely a crucial correction arises from the level density's deformation dependence. Since the level density is generally larger for deformed shapes than for spherical ones, the thermally induced shift of the equilibrium deformation from $\beta^\mathrm{GS}$ (spherical) to $\beta^\mathrm{eq}$ (deformed) enhances the phase space available for neutron emission. This introduces a compensating factor $F$ that mitigates the hindrance of $\Gamma_{n}/\Gamma_{f}$. To account for this, we generalize Eq.~\eqref{eq7gngf} by evaluating the daughter nucleus density at the new equilibrium $\beta^\mathrm{eq}$:

\begin{equation}
\frac{\Gamma_{n}}{\Gamma_{f}}(U^*) \propto \frac{\rho(A-1,\beta^\mathrm{eq}, E_{0} - E_\mathrm{pot}(A-1,U^*) - B_{n})}{\rho(A,\beta^\mathrm{SP}, E_{0} - E_\mathrm{pot}^\mathrm{SP}(A))}.
\end{equation}
Multiplying and dividing by the level density at the ground state $\beta^\mathrm{GS}$, we obtain the modified width ratio:
\begin{equation}\label{eqEb}
\frac{\Gamma_{n}}{\Gamma_{f}}(U^*) = \left(\frac{\Gamma_{n}}{\Gamma_{f}}\right)_0 e^{-\delta E_{\text{mic}}(A-1,U^*)/T} \cdot F,
\end{equation}
where $F = \rho(A-1,\beta^\mathrm{eq}) / \rho(A-1,\beta^\mathrm{GS})$. In deriving this factor, we neglect the minor energy difference between the potentials at $\beta^\mathrm{GS}$ and $\beta^\mathrm{eq}$, as these represent competing minima with comparable energies. Since level densities are significantly higher for deformed configurations ($F > 1$), this factor partially restores the survival probability against the exponential suppression caused by shell damping. The degeneracy effect persists up to $\beta \approx 0.15$ (see Fig.~\ref{Betadependence}).
Consequently, a considerable compensatory effect from Eq.~\eqref{eqEb} is expected when the potential energy minimum shifts from a spherical shape to an equilibrium deformation of $\beta^\mathrm{eq} \approx 0.15$. Increasing the level density parameter in the Fermi gas model by 5$\%$ for $\beta^\mathrm{eq} \approx 0.15$ relative to the spherical GS, we find that $F$ is approximately 5-10.

\paragraph{Conclusions.---}
Our analysis identifies a fundamental mechanism governing the stability of SHN: the \textit{differential thermal damping} of shell corrections. We demonstrate that the shape-dependent level density strictly dictates the nuclear response to excitation. Spherical configurations, while stabilized by deep shell gaps at zero temperature due to high spectral degeneracy, are remarkably fragile against heating. Excitation energy lifts this degeneracy, causing spherical shell corrections to decay faster than those of deformed systems, where symmetry breaking has already spread the levels.

This differential damping drives a thermally induced inversion of the potential energy landscape. As the deep spherical "pocket" vanishes rapidly with temperature, the more persistent deformed minima (oblate or triaxial) become energetically favored. Given the flat macroscopic potential characteristic of SHN, this washing out of spherical effects inevitably shifts the equilibrium toward deformed configurations.

This phenomenon reflects a universal principle of quantum many-body physics: high spectral degeneracy is inherently unstable against symmetry-breaking perturbations. Analogous to the Jahn-Teller effect in molecules, where electronic degeneracy triggers spontaneous distortion, we identify a thermal instability in superheavy nuclei. While the high degeneracy of the spherical potential provides essential stabilization at zero temperature, it renders the system remarkably fragile against heating. Excitation energy drives the nucleus away from the degenerate sphere toward deformed configurations where the lower spectral density ensures more robust shell correlations. Thus, the excitation leads to an averaging of the damping parameter which makes it almost independent of $\beta_{20}^\mathrm{GS}$.

This phenomenon has profound implications for the synthesis of new elements. It implies that hot compound nuclei do not survive as spheres, as traditionally assumed, but equilibrate in deformed shapes. This insight necessitates revising survival probability models, suggesting that while the fission barrier against spherical decay collapses, the emergence of deformed minima may open new, favorable pathways for evaporation-residue formation.

\bibliographystyle{apsrev4-2}
\bibliography{shapes}

\end{document}